\documentstyle[12pt,fleqn]{article}
\def\e{\mbox{e}}
\def\ib{\,\mbox{i}\,}

\def\la{\lambda}

\def\te{\vartheta_1}
\def\tv{\vartheta_4}
\def\case#1#2{{\textstyle{#1\over #2}}}
\def\W#1#2#3#4#5{W #1 \! \left(\hspace{-1mm}
         \begin{array}{cc}#5 & #4 \\ #2 & #3 \end{array}
         \hspace{-1mm}\right)}

\begin{document}

\title{Magnetic Correlation Length and Universal Amplitude of the
Lattice E$_8$ Ising Model}

\author{M.~T.~Batchelor\thanks{
Department of Mathematics, School of Mathematical Sciences,
The Australian National University, 
Canberra, ACT 0200, Australia, e-mail {\tt murrayb@maths.anu.edu.au}}
\hspace{0.2mm} and K.~A.~Seaton\thanks{
Centre for Mathematics and Its Applications,
The Australian National University, 
Canberra, ACT 0200, Australia,
e-mail {\tt k.seaton@latrobe.edu.au}}
\thanks{On leave from School of Mathematics, La Trobe University,
		 Bundoora, Victoria 3083, Australia.}}

\date{ANU-MRR-016-97, May 8, 1997}
\maketitle

\begin{abstract}
The perturbation approach is used to derive the exact 
correlation length $\xi$ of the dilute A$_L$ lattice models in regimes
1 and 2 for $L$ odd. In regime 2
the A$_3$ model is the E$_8$ lattice realisation of the 
two-dimensional Ising model in a magnetic field $h$ at $T=T_c$.
When combined with the singular part $f_s$ of the free energy
the result for the A$_3$ model gives the universal amplitude 
$f_s \,\xi^2 = 0.061~728\ldots$  as $h\to 0$
in precise agreement with the result obtained by 
Delfino and Mussardo via the form-factor bootstrap approach.

\end{abstract}

\newlength{\mathin}
\setlength{\mathin}{\mathindent}
\vskip 1cm

The integrable E$_8$ quantum field theory of Zamolodchikov \cite{Za,Zb} 
is known to be in the same universality class as the two-dimensional 
Ising model in a magnetic field at $T=T_c$.  
Moreover, an integrable lattice realisation of the E$_8$ Ising model
is provided by the dilute A$_3$ model \cite{WNSa,WNSb}, upon which
explicit exact and numerical calculations pertaining to the Ising model 
in a magnetic field can be performed [3-13]. 

In this letter we present the correlation length of the
dilute A$_L$ lattice models in regimes 1 and 2 for $L$ odd, for which the
off-critical perturbation is magnetic-like. This includes the
magnetic correlation length for $L=3$,
of relevance to the magnetic Ising model at $T=T_c$.

The dilute A$_L$ model
is an exactly solvable, restricted solid-on-solid
model defined on the square lattice.
Each site of the lattice can take one of $L$ possible
(height) values, subject to the restriction that
neighbouring sites of the lattice either have the
same height, or differ by $\pm 1$.
The Boltzmann weights of the allowed height configurations of
an elementary face of the lattice are \cite{WNSa,WNSb}
\setlength{\mathindent}{0 cm}
\begin{eqnarray}
\lefteqn{\W{}{a}{a}{a}{a}=
\frac{\te(6\la-u)\te(3\la+u)}{\te(6\la)\te(3\la)}}
\nonumber \\  & & \nonumber \\
\lefteqn{\hphantom{\W{}{a}{a}{a}{a}}
-\left(\frac{S(a+1)}{S(a)}\frac{\tv(2a\la-5\la)}{\tv(2a\la+\la)}
      +\frac{S(a-1)}{S(a)}\frac{\tv(2a\la+5\la)}{\tv(2a\la-\la)}\right)
\frac{\te(u)\te(3\la-u)}{\te(6\la)\te(3\la)}}
\nonumber \\ & & \nonumber \\
\lefteqn{\W{}{a}{a}{a}{a\pm 1}=\W{}{a}{a\pm 1}{a}{a}=
\frac{\te(3\la-u)\tv(\pm 2a\la+\la-u)}{\te(3\la)\tv(\pm 2a\la+\la)}}
\nonumber \\ & & \nonumber \\
\lefteqn{\W{}{a\pm 1}{a}{a}{a}=\W{}{a}{a}{a\pm 1}{a}=
\left(\frac{S(a\pm 1)}{S(a)}\right)^{1/2}
\frac{\te(u)\tv(\pm 2a\la-2\la+u)}{\te(3\la)\tv(\pm 2a\la+\la)}}
\nonumber \\ & & \nonumber \\
\lefteqn{\W{}{a}{a\pm 1}{a\pm 1}{a}=\W{}{a}{a}{a\pm 1}{a\pm 1}}
\nonumber \\ & & \nonumber \\
\lefteqn{ \hphantom{\W{}{a}{a\pm 1}{a\pm 1}{a}}
=\left(\frac{\tv(\pm 2a\la+3\la)\tv(\pm 2a\la-\la)}
           {\tv^2(\pm 2a\la+\la)}\right)^{1/2}
\frac{\te(u)\te(3\la-u)}{\te(2\la)\te(3\la)} }
\nonumber \\ & & \label{Bweights} \\
\lefteqn{\W{}{a}{a\mp 1}{a}{a\pm 1}=
\frac{\te(2\la-u)\te(3\la-u)}{\te(2\la)\te(3\la)}}
\nonumber \\ & & \nonumber \\
\lefteqn{\W{}{a\pm 1}{a}{a\mp 1}{a}=
-\left(\frac{S(a-1)S(a+1)}{S^2(a)}\right)^{1/2}
\frac{\te(u)\te(\la-u)}{\te(2\la)\te(3\la)}}
\nonumber \\ & & \nonumber \\
\lefteqn{\W{}{a\pm 1}{a}{a\pm 1}{a}=
\frac{\te(3\la-u)\te(\pm 4a\la+2\la+u)}{\te(3\la)\te(\pm 4a\la+2\la)}
+\frac{S(a\pm 1)}{S(a)}
\frac{\te(u)\te(\pm 4a\la-\la+u)}{\te(3\la) \te(\pm 4a\la+2\la)}}
\nonumber \\ & & \nonumber \\
\lefteqn{\hphantom{\W{}{a\pm 1}{a}{a\pm 1}{a}}=
\frac{\te(3\la+u)\te(\pm 4a\la-4\la+u)}
{\te(3\la)\te(\pm 4a\la-4\la)}}
\nonumber \\ & & \nonumber \\
\lefteqn{\hphantom{\W{}{a\pm 1}{a}{a\pm 1}{a}}+
\left(\frac{S(a\mp 1)}{S(a)}\frac{\te(4\la)}{\te(2\la)}
-\frac{\tv(\pm 2a\la-5\la)}{\tv(\pm 2a\la+\la)} \right)
\frac{\te(u)\te(\pm 4a\la-\la+u)}{\te(3\la) \te(\pm 4a\la-4\la)}.} 
\nonumber 
\end{eqnarray}

The crossing factors $S(a)$ are defined by
\setlength{\mathindent}{\mathin}
\begin{equation}
S(a)  =  (-1)^{\displaystyle a} \;\frac{\vartheta_1({4a\lambda})}{
           \vartheta_4({2a\lambda})}
\end{equation}
and $\vartheta_1({u})$, $\vartheta_4({u})$ are standard elliptic
    theta functions of nome $p$
\begin{eqnarray}
\vartheta_1(u)&=&\vartheta_1(u,p)=2p^{1/4}\sin u\:
  \prod_{n=1}^{\infty} \left(1-2p^{2n}\cos
   2u+p^{4n}\right)\left(1-p^{2n}\right)\label{theta1}  \\
\vartheta_4(u)&=&\vartheta_4(u,p)=\prod_{n=1}^{\infty}\left(
 1-2p^{2n-1}\cos2u+p^{4n-2}\right)\left(1-p^{2n}\right).
  \label{theta4}
\end{eqnarray}

In the above weights the
variable $\lambda$ and the range of the spectral parameter $u$
are given by $0<u< 3\lambda$ with
\begin{equation}
\lambda =   \frac{s}{r} \; \pi
\end{equation}
where $r=4(L+1)$ and $s=L$ in regime 1 and $s=L+2$ 
in regime 2.\footnote{The model has other regimes, but they are
not of interest here.}  
The magnetic Ising point occurs in regime 2 with $\lambda=\case{5\pi}{16}$.

The row transfer matrix of the dilute A models is
defined on a periodic strip of width $N$ as 
\begin{equation}
T_{\{a\}}^{\{b\}} = \prod_{j=1}^{N}
\W{}{a_j}{a_{j+1}}{b_{j+1}}{b_j} 
\end{equation}
where $\{a\}$ is an admissible path of heights and
$a_{N+1} =a_1$, $b_{N+1} = b_1$. For convenience we
take $N$ even.

The eigenvalues of the transfer matrix are \cite{BNW,ZPG,Zh}
\begin{eqnarray}
\Lambda(u) &=& \omega \left[
\frac{\te(2\lambda-u)\;\te(3\lambda-u)}{\te(2\lambda)\;\te(3\lambda)}
\right]^N
\prod_{j=1}^N
\frac{\te(u-u_j+\lambda)}{\te(u-u_j-\lambda)}
\nonumber \\
&+& \left[
\frac{\te(u)\;\te(3\lambda-u)}{\te(2\lambda)\;\te(3\lambda)}
\right]^N
\prod_{j=1}^N
\frac{\te(u-u_j) \; \te(u-u_j-3\lambda)}
     {\te(u-u_j-\lambda) \; \te(u-u_j-2 \lambda)} 
\label{eigs}
\\
&+& \omega^{-1}
\left[
\frac{\te(u)\;\te(\lambda-u)}{\te(2\lambda)\;\te(3\lambda)}
\right]^N
\prod_{j=1}^N
\frac{\te(u-u_j-4\lambda)}{\te(u-u_j-2\lambda)} \nonumber
\end{eqnarray}
where the $N$ roots $u_j$ are given by the Bethe equations 
\begin{equation}
\omega \left[
\frac{\te(\lambda-u_j)}{\te(\lambda+u_j)}\right]^{N} =
-\prod_{k=1}^{N}
\frac{\te(u_j - u_k - 2\lambda) \; \te(u_j - u_k + \lambda) }
     {\te(u_j - u_k + 2\lambda) \; \te(u_j - u_k - \lambda) }
\label{BAE}
\end{equation}
and $\omega=\exp(\ib \pi \ell/(L+1))$ for $\ell=1,\ldots,L$.

There are several methods at hand to calculate the correlation length.
Here we apply the perturbative approach initiated by 
Baxter \cite{Baxter,PB}.
For $L$ odd this involves perturbing away from the strong magnetic 
field limit at $p=1$. We thus introduce the variables
\begin{equation}
w = \e^{-2\pi u/\epsilon} \quad \mbox{and} \quad 
x = \e^{- \pi^2/r \epsilon}
\end{equation}  
conjugate to the nome $p=\e^{-\epsilon}$.
The relevant conjugate modulus transformations are
\begin{eqnarray}
\te(u,p) &=& \left( {\pi \over \epsilon} \right)^{1/2} 
             \e^{-(u-\pi/2)^2/\epsilon} \,
              E(w,q^2) \\
\tv(u,p) &=& \left( {\pi \over \epsilon} \right)^{1/2} 
             \e^{-(u-\pi/2)^2/\epsilon} \,
              E(-w,q^2) 
\end{eqnarray}
where $q=\e^{-\pi^2/\epsilon}$ and 
\begin{equation}
E(z,p) = \prod_{n=1}^{\infty} (1-p^{n-1} z)(1-p^n z^{-1})(1-p^n).
\end{equation}

In the ordered limit ($p\to 1$ with $u/\epsilon$ fixed) the 
Boltzmann weights for $L$ odd reduce to
\begin{equation}
\W{}{a}{b}{c}{d} \sim w^{H(d,a,b)} \, \delta_{a,c} .
\end{equation}
The function $H(d,a,b)$ is given explicity in \cite{WPSN},
being required for the calculation of the local height
probabilities. In this limit the row transfer matrix 
eigenspectra breaks up into a number of bands labelled by
integer powers of $w$. In regime 1 there are $\case{1}{2}(L+1)$ 
ground states and in regime 2 there are
$\case{1}{2}(L-1)$ ground states, each with eigenvalue $\Lambda_0=1$. 
The bands of excitations are relevant to the calculation of the
correlation length. 

The number of states in the $w$ band is $\case{1}{2}(L-1) N$ in
regime 1 and $\case{1}{2}(L-3) N$ in regime 2. These correspond to
introducing in all but one of the ground state paths $\{a\}$
a single non-ground state height, in any position. 
In particular, note that there are {\em no} excitations in
the $w$ band for $L=3$ in regime 2. Thus for the magnetic Ising model
we must consider excitations in the $w^2$ band. These are harder to
count, arising from a variety of both single and multiple 
deviations from ground state paths.
However, we observe numerically that (apart from when $N=2$) 
there are $4N$ states in the $w^2$ band.    

We associate a given value of $\ell$ with each eigenvalue
by numerically comparing the eigenspectrum at criticality ($p=0$) 
with the eigenspectrum of the corresponding O($n$) loop model \cite{WN}
for finite $N$.\footnote{Strictly speaking we compare with the
eigenspectrum of the corresponding vertex model with seam $\omega$.} 
Each eigenvalue can then be tracked to the ordered limit. The band
of largest eigenvalues is seen to have the values
$\ell = 1,\ldots,\case{1}{2}(L+1)$ in regime 1 and
$\ell = 1,\ldots,\case{1}{2}(L-1)$ in regime 2.
 
Setting $w_j = \e^{-2\pi u_j/\epsilon}$, the eigenvalues (\ref{eigs}) 
can be written
\setlength{\mathindent}{0 cm}
\begin{eqnarray}
\Lambda(w) &=& \omega \left[
\frac{E(x^{4s}/w,x^{2r})\;E(x^{6s}/w,x^{2r})}
     {E(x^{4s},x^{2r})\;E(x^{6s},x^{2r})}
\right]^N
\prod_{j=1}^N w_j^{1 - 2s/r} \,
\frac{E(x^{2s} w /w_j,x^{2r})}{E(x^{2s} w_j/w,x^{2r})}
\nonumber \\
&+& \left[ 
\frac{x^{2s}}{w}
\frac{E(w,x^{2r})\;E(x^{6s}/w,x^{2r})}
     {E(x^{4s},x^{2r})\;E(x^{6s},x^{2r})}
\right]^N
\prod_{j=1}^N w_j \,
\frac{E(w/w_j,x^{2r})\;E(x^{6s} w_j /w,x^{2r})}
     {E(x^{2s} w_j/w,x^{2r})\;E(x^{4s} w_j /w,x^{2r})}
\nonumber\\
&+& \omega^{-1}
\left[ x^{2s} \,
\frac{E(w,x^{2r})\;E(x^{2s}/w,x^{2r})}
     {E(x^{4s},x^{2r})\;E(x^{6s},x^{2r})}
\right]^N
\prod_{j=1}^N w_j^{2s/r}
\frac{E(x^{8s} w_j /w,x^{2r})}{E(x^{4s} w_j /w,x^{2r})} .
\end{eqnarray}
The Bethe equations (\ref{BAE}) are now 
\setlength{\mathindent}{\mathin}
\begin{equation}
\omega \left[ w_j \, 
\frac{E(x^{2s}/w_j,x^{2r})}{E(x^{2s} w_j,x^{2r})}
\right]^N
= - \prod_{k=1}^N w_k^{2s/r}
\frac{E(x^{2s} w_j/w_k,x^{2r}) \, E(x^{4s} w_k/w_j,x^{2r})}
     {E(x^{2s} w_k/w_j,x^{2r}) \, E(x^{4s} w_j/w_k,x^{2r})} .
\end{equation}

The calculation of the largest eigenvalue proceeds from the
$x\to0$ limit with $w$ fixed in a similar manner
to that for the eight-vertex \cite{Baxter} and CSOS \cite{PB} models. 
Each of the degenerate ground states has a different root
distribution $\{w_j\}$ on the unit circle, depending on $\ell$. 
Defining the free energy per site as $f = N^{-1} \log \Lambda_0$
our final result is
\setlength{\mathindent}{0 cm}
\begin{equation}
f =  4 \sum_{k=1}^{\infty}
\frac{ \cosh [(5 \lambda-\pi) \pi k/\epsilon] \,
\cosh (\pi \lambda k/\epsilon) \,
\sinh (\pi u k/\epsilon) \,
\sinh [(3\lambda-u)\pi k/\epsilon]}
{k\, \sinh (\pi^2 k/\epsilon) \,
\cosh (3 \pi \lambda k/\epsilon)} \label{fen}
\end{equation}
in agreement with the previous calculations via the inversion
relation method \cite{WNSa,WNSb,WPSN}.

In regime 1, the leading eigenvalue in the $w$ band has 
$\ell=\case{1}{2}(L+1)+1$. The root distribution has
$N-1$ roots on the unit circle and a 1-string excitation
located exactly at $w_N = - x^r$. 
Applying perturbative arguments along the lines of \cite{PB} 
yields the leading excitation in the $w$ band to be
\setlength{\mathindent}{\mathin}
\begin{equation}
\frac{\Lambda_1}{\Lambda_0} = w\, 
\frac{E(-x^{2s}/w,x^{12s})\,E(-x^{4s}/w,x^{12s})}
     {E(-x^{2s}\,w,x^{12s})\,E(-x^{4s}\,w,x^{12s})}.
\label{reg1}
\end{equation}
At the isotropic point $w = x^{3s}$ this reduces to
\begin{equation}
\frac{\Lambda_1}{\Lambda_0} = x^{s}\,
\frac{E^2(-x^s,x^{12s})}
     {E^2(-x^{5s},x^{12s})} =
\left[
\frac{\tv(\frac{\pi}{12},p^{\pi/6\lambda})}
     {\tv(\frac{5\pi}{12},p^{\pi/6\lambda})}
\right]^2 .
\end{equation}

For $L=3$ in regime 2 extensive numerical investigations of the 
Bethe equations have lead to a convincing conjecture for the
thermodynamically significant strings \cite{BNW,GN}.
We find that the leading excitation in the $w^2$ band
is a 2-string with $\ell=2$. However, the state is originally a
1-string for small $p$. Such behaviour has been discussed
in \cite{GN}. Tracking this state with increasing $p$ reveals that
the 2-string is exactly located at $-x^{\pm 11}$ in the limit $p=1$.
There are finite-size deviations away from this position for
small $N$ and $0 < p < 1$. The location we find for this string is in
accord with the previous numerical work \cite{BNW,GN}.  
Applying the perturbation arguments in this case yields the
leading excitation in the $w^2$ band for $L=3$ to be
\begin{equation}
\frac{\Lambda_2}{\Lambda_0} = w^2\,
\frac{E(-x/w,x^{60})\,E(-x^{11}/w,x^{60})\,
      E(-x^{31}\,w,x^{60})\,E(-x^{41}\,w,x^{60})}
     {E(-x\,w,x^{60})\,E(-x^{11}\,w,x^{60})\,
      E(-x^{31}/w,x^{60})\,E(-x^{41}/w,x^{60})} .
\label{reg2}
\end{equation}
At the isotropic point $w = x^{15}$ this reduces to
\begin{equation}
\frac{\Lambda_2}{\Lambda_0} = x^{28}\,
\frac{E^2(-x^4,x^{60})\,E^2(-x^{14},x^{60})}
     {E^2(-x^{16},x^{60})\,E^2(-x^{26},x^{60})} = 
\left[
\frac{\tv(\frac{\pi}{15},p^{8/15})\,\tv(\frac{7\pi}{30},p^{8/15})}
     {\tv(\frac{4\pi}{15},p^{8/15})\,\tv(\frac{13\pi}{30},p^{8/15})} 
\right]^2 .
\end{equation}

The correlation length $\xi$ can be obtained either by integrating over
the relevant band of eigenvalues or via the leading
eigenvalue in the band at the isotropic point (see, e.g., \cite{PB}). 
Doing the latter we have   
\begin{equation}
\xi^{-1} = - \log \frac{\Lambda}{\Lambda_0}
\end{equation}
where $\Lambda$ is the relevant leading eigenvalue.
Our final results are thus
\begin{equation}
\xi^{-1} = 2 \log \left[
\frac{\tv(\frac{5\pi}{12},p^{\pi/6\lambda})}
     {\tv(\frac{\pi}{12},p^{\pi/6\lambda})}   
\right] \label{cor1}
\end{equation}
for $L$ odd in regime 1, with 
\begin{equation}
\xi^{-1} = 2 \log \left[
\frac{\tv(\frac{4\pi}{15},p^{8/15})\,\tv(\frac{13\pi}{30},p^{8/15})}
{\tv(\frac{\pi}{15},p^{8/15})\,\tv(\frac{7\pi}{30},p^{8/15})}
\right] \label{cor2}
\end{equation}
for $L=3$ in regime 2.

The derivation of the correlation length for $L\ne3$ in regime 2
is complicated. In this regime the leading excitation in the $w$
band has $\ell=\case{1}{2}(L-1)+1$ and, like the leading 2-string
in the $w^2$ band for $L=3$, it begins life for small $N$ and
$p \simeq0$ as a 1-string. We have not pursued this further. 
Nevertheless we have numerically
observed that the final result (\ref{reg1}) also applies to
the leading $w$ band excitation in regime 2. We thus believe
that the correlation length (\ref{cor1}) and the corresponding
exponents below also hold in regime 2 for $L \ne 3$.

It follows from (\ref{cor1}) that the correlation length
diverges at criticality as 
\begin{equation}
\xi \sim {1 \over 4 \sqrt 3}\, p^{-\nu_h} \quad \mbox{as} \quad p \to 0  
\end{equation}
where the correlation length exponent $\nu_h$ is given by  
\begin{equation}
\nu_h = \frac{r}{6 s} = 
\left\{
\begin{array}{lll}
\frac{2(L+1)}{3 L}   & \qquad & \mbox{regime 1} \\
& & \\ 
\frac{2(L+1)}{3(L+2)}   & \qquad & \mbox{regime 2 }. 
\end{array}
\right.  \label{exps}
\end{equation}

The correlation length exponents are seen to satisfy the 
general scaling relation 
$2 \nu_h = 1 + 1/\delta$, which follows from the relation 
\begin{equation}
f_s \, \xi^2 \sim \mbox{constant}
\end{equation}
where $f_s \sim p^{1+ 1/\delta}$ is the singular part of the bulk 
free energy and the exponents $\delta$ are those following from 
the singular behaviour of (\ref{fen}) 
\cite{WNSa,WNSb,WPSN}.\footnote{The same correlation length exponents
should hold for $L$ even, for which the integrable perturbation is 
thermal-like. The scaling relation
is now $2\nu_t = 2 - \alpha$, where $\nu_t$ and $\alpha$ are as given
in (\ref{exps}) and \cite{WNSa,WNSb,WPSN}, respectively. In particular,
(\ref{exps}) gives the Ising value $\nu_t=1$ for $L=2$ in regime 1,
as expected.}

The magnetic Ising case at $\lambda=\case{5\pi}{16}$ is of particular 
interest. From (\ref{fen}) we find
\begin{equation}
f_s \sim 4 \sqrt 3 \, \frac{\sin{\pi \over 5}}{\cos{\pi \over 30}} 
                   \, p^{16/15}
\quad \mbox{as} \quad p \to 0 \, .
\end{equation}
On the other hand, from (\ref{cor2}) we have
\begin{equation}
\xi \sim {1 \over 8 \sqrt 3 \, \sin{\pi \over 5}} \, p^{-8/15} 
\quad \mbox{as} \quad p \to 0 \, .
\end{equation}
Combining these results gives the universal magnetic Ising amplitude 
\begin{equation}
f_s \, \xi^2 = {1 \over 16 \sqrt 3 \sin{\pi \over 5} \cos{\pi \over 30}} 
             = 0.061~728~589 \ldots \quad \mbox{as} \quad p \to 0 \, .
\end{equation} 
This is in precise agreement with the field-theoretic result obtained 
recently by Delfino and Mussardo, starting from
Zamolodchikov's $S$-matrix and using the 
form-factor bootstrap approach \cite{F,DM}.
Full details of our calculations will be given elsewhere.

\vskip 5mm
It is a pleasure to thank John Cardy and Ole Warnaar for some
helpful remarks. The work of KAS has been facilitated by a Commonwealth
Staff Development Fund grant, administered by the Academic Development
Unit of La Trobe University.
The work of MTB has been supported by the Australian Research Council.


\end{document}